# Complete Coherent Control of Silicon-Vacancies in Diamond Nanopillars Containing Single Defect Centers

Jingyuan Linda Zhang[‡,1], Konstantinos G. Lagoudakis[‡,1], Yan-Kai Tzeng[4], Constantin Dory[1], Marina Radulaski[1], Yousif Kelaita[1], Kevin A. Fischer[1], Zhi-Xun Shen[2,3,4], Nicholas A. Melosh[2,3], Steven Chu[4,5], Jelena Vučković[1]

[1]E. L. Ginzton Laboratory, Stanford University, Stanford, California 94305, USA
[2]Geballe Laboratory for Advanced Materials, Stanford University, Stanford, California 94305, United States
[3]Stanford Institute for Materials and Energy Sciences, SLAC National Accelerator Laboratory, Menlo Park, California 94025, USA
[4]Department of Physics, Stanford University, Stanford, California 94305, USA
[5]Department of Molecular and Cellular Physiology, Stanford University, Stanford, California 94305, USA
[‡]These authors contributed equally.

**Correspondence:**
J. L. Zhang, e-mail: *ljzhang@stanford.edu* ; K. G. Lagoudakis, e-mail: *lagous@stanford.edu*

**Abstract**

**Arrays of identical and individually addressable qubits lay the foundation for the creation of scalable quantum hardware such as quantum processors and repeaters. Silicon vacancy centers in diamond ($SiV^-$) offer excellent physical properties such as low inhomogeneous broadening, fast photon emission, and a large Debye-Waller factor, while the possibility for all-optical ultrafast manipulation and techniques to extend the spin coherence times make them very promising candidates for qubits. Here, we have developed arrays of nanopillars containing single $SiV^-$ centers with high yield, and we demonstrate ultrafast all-optical complete coherent control of the state of a single $SiV^-$ center. The high quality of the chemical vapor deposition (CVD) grown $SiV^-$ centers provides excellent spectral stability, which allows us to coherently manipulate and quasi-resonantly read out the state of individual $SiV^-$ centers on picosecond timescales using ultrafast optical pulses. This work opens new opportunities towards the creation of a scalable on-chip diamond platform for quantum information processing and scalable nanophotonics applications.**

**Introduction**

Deterministic positioning of individual quantum emitters with nearly identical properties is an outstanding challenge towards the creation of large scale quantum hardware. Recent technological leaps have enabled the development of high quality site-controlled quantum dots that have attracted a strong interest for their potential use as building blocks in scalable quantum hardware[1-3], yet the inhomogeneous broadening of such emitters still imposes limitations towards large scale integration. An alternative system that has arisen as a promising candidate for scalable qubits is the nitrogen vacancy (NV$^-$) centers in diamond, due to its excellent spin properties under ambient conditions[4-6] and potential for on-demand positioning[7,8]. However, spectral diffusion and its large phonon sideband limit both the quality of coherent photon generation from NV$^-$ centers and their ability to be integrated in devices that might have nearby free surfaces.

Most recently, SiV$^-$ centers in diamond have attracted attention as a potential candidate for qubits. Due to the inversion symmetry of the color center[9], SiV$^-$s offer spectral stability[10], strong emission into the zero phonon line[11], and narrow inhomogeneous broadening[9,12]. Remarkable efforts have been made to yield nanophotonic devices in bulk diamond, for integration of these color centers into a photonic network[13-19] and to deterministically position them[20]. In the meantime, the spin coherence time has been measured, and both techniques to extend the spin coherence time and fast spin manipulation are under active investigation[21-23].

Here we fabricated regular arrays of diamond nanopillars containing individual SiV$^-$ centers with high yield and spectral stability, using diamond grown by chemical vapor deposition (CVD) followed by electron-beam lithography and dry etching to generate the nanophotonic structures. Taking advantage of the exceptional properties of CVD grown SiV$^-$ centers in diamond, we use this extremely promising platform to demonstrate for the first time complete SU(2) coherent control of single SiV$^-$ centers in nanopillars using ultrafast optical pulses, enabled by the high post-fabrication quality of the SiV$^-$ centers. The ability to coherently control an isolated low strain SiV$^-$ center on a picosecond timescale, in a photonic device tens of nanometers from the diamond surface, paves the way towards scalable on-chip integration of these diamond-defect centers into nanophotonic devices for highly scalable quantum hardware.

**Materials and Methods**

The sample is grown by microwave plasma chemical vapor deposition (MPCVD) method on a high purity type IIa diamond substrate from Element Six. Nominally, a 100-nm-thick layer of diamond containing SiV⁻ is grown homoepitaxially on the diamond substrate, with the following growth conditions: $H_2$: 300 sccm, $CH_4$: 0.5 sccm, stage temperature 650 °C, microwave power 1.3 kW, and pressure: 23 Torr. The silicon atoms are generated from the hydrogen plasma etching of the silicon wafer, and SiV⁻ centers are subsequently formed through silicon incorporated into the grown layer by plasma diffusion. Following growth, a pattern of 8 by 8 arrays of diamond nanopillars with interleaved crossbars is exposed in a hydrogen silsesquioxane (HSQ) mask using electron-beam lithography (the crossbars are added as markers to facilitate locating the nanopillars under white light imaging). Lastly, an oxygen-based reactive ion etching (RIE) transfers the pattern into the diamond, and the HSQ mask is removed in hydrofluoric acid, resulting in bare diamond pillars. The bare nanopillars are 135-170 nm in diameter and 200 nm in height. Scanning electron microscope (SEM) images of a nanopillar array and a single nanopillar are shown in Figure 1a and Figure 1b, respectively. In addition to the arrays, we have incorporated large square areas on the same sample to produce a minimally strained environment that yields an ensemble of SiV⁻ centers comparable to those in bulk diamond.

The photoluminescence (PL) spectrum at 10 K of the single SiV⁻ in nanopillar under investigation is presented in Figure 1c (red), superposed with the PL from an ensemble of SiV⁻s in bulk diamond (black). The SiV⁻ centers in the nanopillars show low spectral shifting compared to those in low-strain bulk diamond, with the energy of each transition lying within the inhomogeneous distribution of the SiV⁻ ensemble in bulk. The height of the nanopillars is greater than the thickness of the epilayer containing the SiV⁻s, such that after the etching, the SiV⁻ centers are confined in the nanostructures masked by the negative tone resist. This is confirmed by scanning confocal microscopy of the sample surface, as shown in Figure 1d. The custom-made laser scanning confocal microscope consists of 532 nm CW laser focused onto the sample surface through a high numerical aperture (NA = 0.75) microscope objective, while the photoluminescence (PL) from the excited spot is collected through the same objective, filtered by a dichroic and a long pass filter, into a single mode fiber. The excitation/collection spot is scanned across the sample surface with a fast steering mirror. The PL signal is sent either to a single photon counting module (SPCM) for construction of the PL map (Figure 1d) or to a high-resolution spectrometer for spectroscopic analysis (Figure 1c). In the PL map of a representative portion of a nanopillar array on a higher SiV¯ density sample in Figure 1d, the crossbars and nanopillars containing SiV¯ centers produce high photon count rates under 532 nm laser excitation, and the photoluminescence (PL) spectra from these areas confirm the presence of SiV¯ centers while the spectra from the background areas show no SiV¯ emission (not shown).

The fabricated nanopillar arrays contain isolated single SiV¯ centers with high yield. A statistical study of the PL spectra was conducted on several arrays, one of which is presented in the Supplementary Information. In the case of more than one emitter per nanopillar, due to the inhomogeneous distribution and in some cases the presence of strain, the spectra contain distinguishable sets of four transitions associated with each SiV¯ center. Therefore, by counting the number of narrow lines in each PL spectrum, one can estimate the number of color centers in each nanopillar. Out of the 64 nanopillars investigated within an entire array, 40, 20 and 4 nanopillars contain 0, 1, and 2 SiV¯ centers per pillar, respectively, very consistent with a Poisson distribution with an average value of 0.44. Second-order autocorrelation measurements of $g^{(2)}(\tau)$ yield $g^{(2)}(0)<0.5$ (see Supplementary Information), corroborating the single emitter nature of the SiV¯ in nanopillar devices, and further supports the conclusions from the spectroscopic study. Based on the statistics of the number of SiV¯ centers embedded in the pillars, we deduce the density of SiV¯ centers in the epilayer to be $\sim 3\times 10^{14}$ cm$^{-3}$.

The spectroscopic studies were performed using an above-band excitation laser at 720 nm in combination with a custom-built double grating spectrometer with ~5.0 GHz resolution. Coherent control of the SiV¯ center was performed with spectrally filtered resonant pulses from a mode-locked Ti:Sapphire laser, which allows for addressing individual transitions. The double-grating spectrometer enables the spectral filtering of the emission from individual transitions, which is subsequently collected by a single photon counting module (SPCM) while rejecting light from the resonant spectrally filtered pulses.

**Results and Discussion**

Having identified the pillars with individual SiV¯ centers, we then proceed with the demonstration of all-optical coherent manipulation. At low temperature (10 K), the PL spectrum of the SiV⁻ center under investigation displays four narrow peaks (Figure 2a), corresponding to the four radiative transitions between the two excited states and the two ground states as shown in the inset of Figure 2a. To manipulate the state of the SiV⁻ center, a linearly polarized, spectrally filtered picosecond optical pulse is applied to resonantly drive transition $|4\rangle \leftrightarrow |2\rangle$. The spectral shape of the resonant driving pulse is shown as the blue shaded area in Figure 2a. The pulse coherently rotates the state between the upper ground state $|2\rangle$ and the upper excited state $|4\rangle$, where we define these two levels as our qubit. Because of fast phonon-mediated relaxation processes between the two excited states, as shown by the grey arrow in the inset of Figure 2a, level $|3\rangle$ is populated proportionally to the population of level $|4\rangle$. A modulation of the upper excited state population thereby modulates the fluorescence intensity of transition $|3\rangle \rightarrow |1\rangle$. Using the double spectrometer, we spectrally filter transition $|3\rangle \rightarrow |1\rangle$ which is used as the detection channel (yellow shaded area in Figure 2a). Detection of a photon from this transition indicates that the system has been set to state $|1\rangle$. Due to the short orbital relaxation time of the ground states of the SiV⁻, level $|2\rangle$ is thermally repopulated according to the Boltzmann distribution[24] and a new qubit rotation between $|4\rangle \leftrightarrow |2\rangle$ may occur. Because transition $|3\rangle \rightarrow |1\rangle$ is ~200 GHz away from our resonant pulses, this detection scheme enables us to take full advantage of the double spectrometer and almost entirely suppress leakage of the pulses in our detection channel. Moreover, to the best of our knowledge, this is the first demonstration of coherent control of individual SiV⁻ centers in diamond through direct detection of the SiV⁻ zero phonon line emission, while previous studies of the dynamics of SiV⁻ center have been performed through collection of the phonon sideband[21-23].

The effect of the resonant pulses on the upper excited state population is shown in Figure 2b. The pulses induce Rabi rotations on our qubit that are evidenced by oscillations of the detected photon counts as a function of the pulse area. The measured rotation angle extends to $3\pi$, as shown in Figure 2b, with the solid red curve representing a sinusoidal fit with a linear background proportional to the power. On the Bloch sphere (inset of Figure 2b), this corresponds to a rotation of the Bloch vector about the x-axis, bringing the qubit from the north (state $|2\rangle$) to the south pole (state $|4\rangle$) and back.

To demonstrate rotation of the qubit about an arbitrary axis on the Bloch Sphere, we proceeded with the demonstration of Ramsey interference. A Ramsey interference experiment requires two π/2 pulses and a variable interpulse delay τ – the first pulse introduces a θ=π/2 Rabi rotation of the Bloch vector about the x-axis, and the second pulse further rotates the Bloch vector by θ=π/2 about an axis that is at φ=$ω_L$τ from the x-axis, with τ the delay between the two pulses and $ω_L$ the frequency of the driven transition. For φ=2nπ, the second pulse brings the Bloch vector to state |4⟩, resulting in maximum counts, whereas for φ=(2n+1)π the Bloch vector is brought back to state |2⟩ and the resulting detected counts are at a minimum. A continuous variation of the interpulse delay results in a "figure eight" path on the Bloch sphere, with the two tips located at the north and south poles of the sphere and the central cross point along the y-axis on the equator. To avoid interference between the two π/2 pulses, we tune the coarse interpulse delay to τ ≥ 66.7 ps, implemented using a Mach-Zehnder interferometer with a motorized delay stage in one arm for the coarse delay and a piezoelectric actuator in the other arm for the fine delay. Ramsey interference is obtained with the piezoelectric actuator, which introduces fine delays of the order of 10 fs, enough to resolve the oscillations at the optical frequency. An example of the Ramsey interference at a coarse delay of 100.08 ps is shown in the inset of Figure 3a. The oscillations are fitted with a sinusoidal function, from which we extract the contrast of the oscillations for the particular coarse delay. It is important to note that a Ramsey experiment is a quantum interference phenomenon, and therefore it is extremely sensitive to both the frequency stability of the emitter and the phase stability of the set-up – an interpulse delay fluctuation or emitter frequency shift manifests itself as a phase shift in the Ramsey interference. Therefore, the high-quality Ramsey interference in the inset of Figure 3a shows the exceptional stability of our quantum emitters and the experimental apparatus. Repeating this experiment for a series of coarse delays (Figure 3a, red filled circles) allows us to estimate the dephasing time $T_2^*$ of the qubit. In particular, by fitting a Gaussian function to the Ramsey interference contrast decay[25] in Figure 3b, we obtain $T_2^*$ = 240 ps (red line, $\exp(-t^2/T_2^{*2})$) for this particular SiV⁻, in good agreement with previously measured values[21]. This is limited by the excited state lifetime of 1.7 ns at liquid helium temperature[10], further shortened due to the fast decay into the lower excited state and any local strain in the material.

To demonstrate universal single qubit gate operation, we performed complete coherent control which allows for access to the entire Bloch sphere. Full SU(2) control of our qubit is achieved through a slight variation of the Ramsey interference experiment by driving the qubit with dual resonant pulses with both variable area and variable delay. The final state of the qubit after the control pulses is shown in Figure 4a for some combinations of the pulse areas and interpulse delays. As with the Ramsey experiments, we performed the SU(2) control at 66.7 ps coarse delay to avoid interference between the two pulses. After setting the amplitude of both pulses, the fine interpulse delay is scanned over 10 fs while recording the photon counts from the detection channel. The Ramsey-type interference experiment is repeated for all accessible pulse powers, and the resulting photon count map as a function of interpulse fine delay and individual pulse power is provided in Figure 4b, showing the typical multi-lobed structure expected of complete coherent control of resonantly driven qubits[1]. Again, emitter spectral stability and set-up phase stability over several hours of acquisition time were necessary to produce the high-quality SU(2) control results shown in Figure 4b.

**Conclusions**

We have realized a novel platform for scalable quantum information processing based on arrays of diamond nanopillars with an ~150 nm device footprint featuring embedded individually addressable single $SiV^-$ centers with 31.3% yield. Using this platform, we have demonstrated full SU(2) coherent control of individual $SiV^-$ centers using ultrafast optical pulses, which enables universal single qubit gate operation. Moreover, the coherent control of a $SiV^-$ center inside a nanopillar structure is possible as a result of the high spectral stability of these $SiV^-$ centers, despite the less than 75 nm distance from the diamond-air interface. Our demonstration paves the way for future integration of shallow $SiV^-$ centers into nanophotonic devices for cavity quantum electrodynamics (QED) and spin-photon interfaces[26, 27]. Thanks to the low inhomogeneous broadening, it is possible to envision entanglement of two low strain single $SiV^-$ centers hosted in different nanopillars. Finally, the ultrafast all-optical coherent control techniques demonstrated in this work can be extended to the control of Zeeman-split spin states of $SiV^-$ centers. The relatively short electron spin coherence time limited by the single phonon decoherence mechanism can potentially be alleviated in these small nanopillars due to the reduced phonon density of states. Altogether, we show a promising step towards a platform for scalable on-chip quantum systems that could play an important role in future quantum hardware[28].

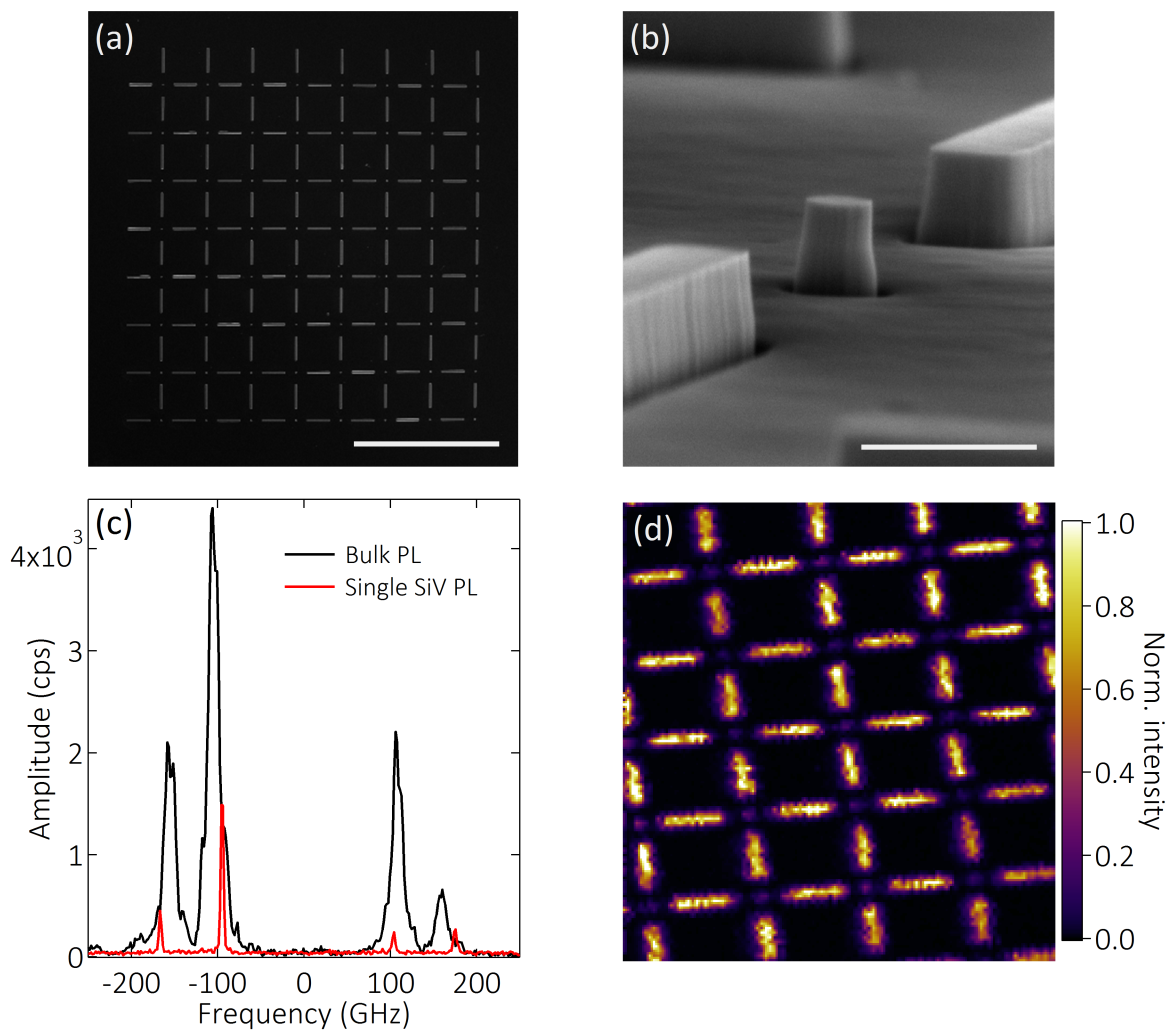

**Figure 1.** (a) Scanning Electron Microscopy (SEM) image of a nanopillar array. Scale bar, 10 μm. (b) SEM image of a 165 nm diameter, 200 nm tall nanopillar. Scale bar, 400 nm. (c) Photoluminescence spectrum from a single nanopillar (red) compared with that from SiV¯ ensemble in bulk diamond (black). (d) Scanning confocal microscopy map of a representative portion of a nanopillar array on a higher SiV¯ center density sample. The bright areas with higher photon count rate correspond to the areas containing SiV¯ centers, while the background is SiV¯ center free.

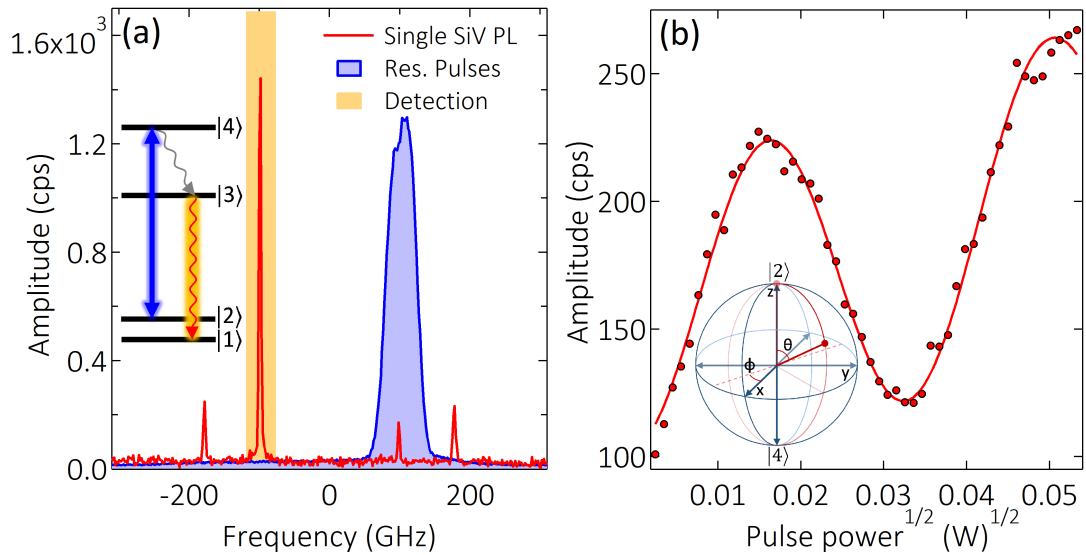

**Figure 2.** (a) Excitation and detection scheme of the coherent control experiment. Transition $|2\rangle \leftrightarrow |4\rangle$ of the single SiV$^-$ center (our qubit) is resonantly addressed with a pulsed Ti:Sapphire laser (blue shaded line), while the radiative emission from transition $|3\rangle \rightarrow |1\rangle$ is detected through a double monochromator by an SPCM (yellow shaded area). The inset shows the energy levels of the SiV$^-$ center. The double-sided thick blue arrow denotes the coherent interaction of the SiV$^-$ with the resonant pulses and the red wavy downward arrow denotes the detected photons coming from transition $|3\rangle \rightarrow |1\rangle$. (b) Demonstration of Rabi rotations. Varying the area of the resonant pulses rotates qubit about the x-axis with a direct impact on the detected photon counts from transition $|3\rangle \rightarrow |1\rangle$, which shows clear oscillations in detected photon counts as a function of the pulse area.

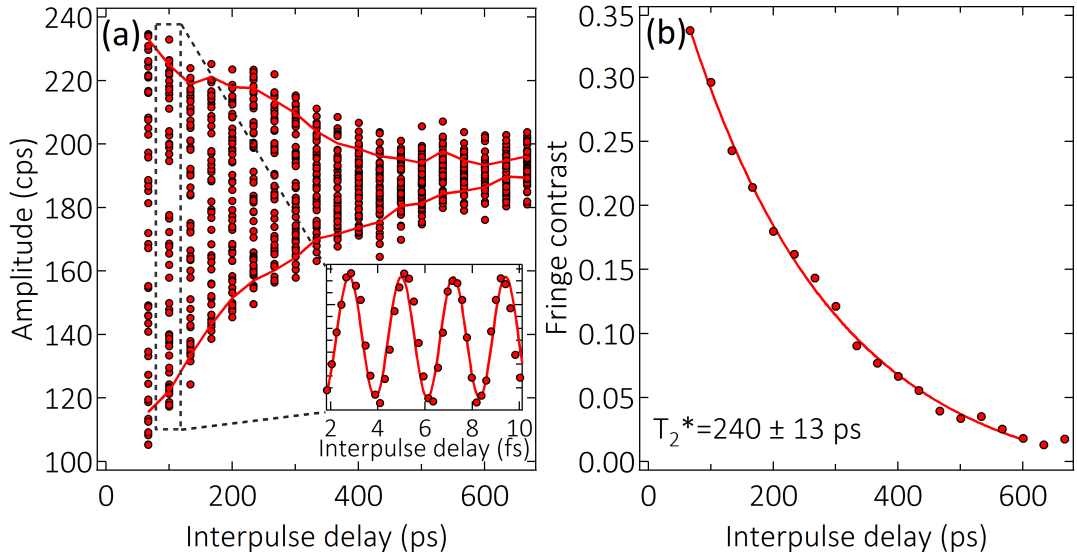

**Figure 3.** (a) Ramsey interference for coarse interpulse delays ranging from 66.7 to 667.2 ps. At each coarse delay, we record the Ramsey interference by varying the fine delay over 10 fs, and then fit it with a sinusoidal function as shown in the inset. The fitted amplitude of the Ramsey interference at each coarse delay is shown by the red envelope. (b) Decay of the Ramsey fringe contrast extracted from the data in (a) (red filled circles). The contrast decay is fitted with a Gaussian function that yields a decay time $T_2^*$ of 240 ps for the qubit.

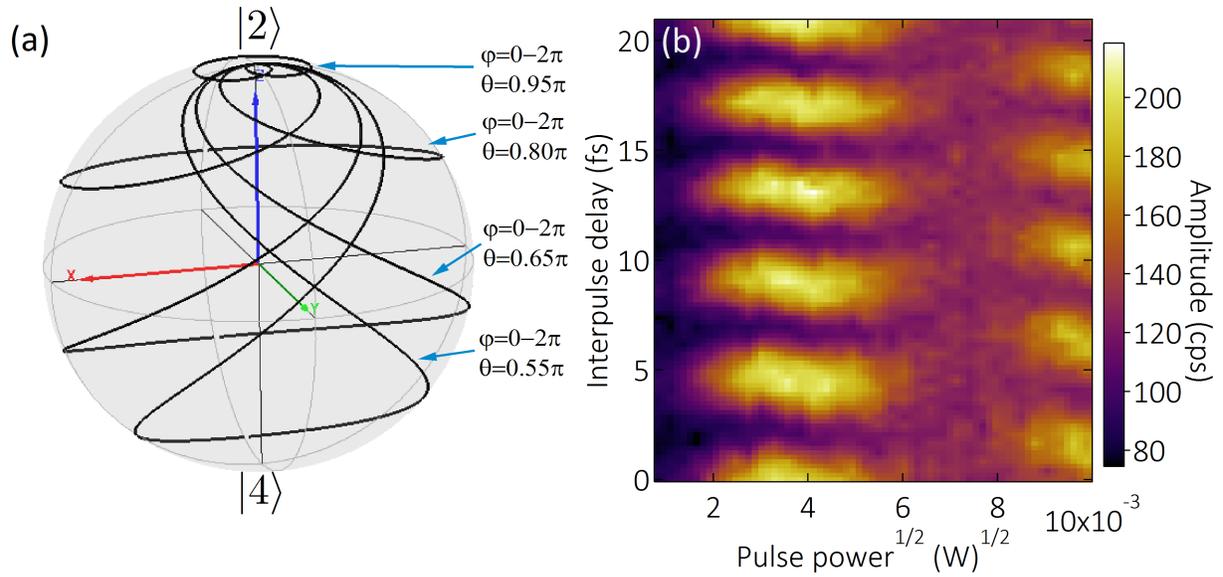

**Figure 4.** Experimental SU(2) control of the upper excited state population. (a) Four distinct closed loops on the Bloch sphere denote the final qubit states after a pair of control pulses with Φ =0–2π interpulse delay for θ=0.95π, 0.80π, 0.65π and 0.55π per pulse, showing that access to the complete sphere is possible through dual pulse excitation with variable amplitudes and interpulse delays. (b) Detected counts for dual pulse excitation with variable pulse area and delay. Here, the angle of rotation per pulse on the Bloch sphere is varied with the pulse area, while the axis about which the state is rotated is controlled with the delay between pulses. The data were taken for a coarse interpulse delay of 66.7 ps.

**Competing Financial Interests**

The authors declare no competing financial interest.

**Acknowledgments**

Financial support for materials synthesis is provided by the Department of Energy Office of Basic Energy Sciences, Division of Materials Sciences through Stanford Institute for Materials and Energy Sciences (SIMES) under contract DE-AC02-76SF00515, and by the Army Research Office under contract W911NF1310309, and in part by the AFOSR MURI for Quantum Metaphotonics and Metamaterials. This work was performed in part at the Stanford Nanofabrication Facility of NNIN supported by the National Science Foundation under Grant No. ECS-9731293, and Stanford Nano Shared Facility. JLZ acknowledges support from the Stanford Graduate Fellowship. CD acknowledges support from the Andreas Bechtolsheim Stanford Graduate Fellowship.

# Supplementary Information

# Complete Coherent Control of Silicon-Vacancies in Diamond Nanopillars Containing Single Defect Centers

Jingyuan Linda Zhang[‡1], Konstantinos G. Lagoudakis[‡1], Yan-Kai Tzeng[4], Constantin Dory[1], Marina Radulaski[1], Yousif Kelaita[1], Kevin A. Fischer[1], Zhi-Xun Shen[2,3,4], Nicholas A. Melosh[2,3], Steven Chu[4,5], Jelena Vučković[1]

[1]E. L. Ginzton Laboratory, Stanford University, Stanford, California 94305, USA
[2]Geballe Laboratory for Advanced Materials, Stanford University, Stanford, California 94305, United States
[3]Stanford Institute for Materials and Energy Sciences, SLAC National Accelerator Laboratory, Menlo Park, California 94025, USA
[4]Department of Physics, Stanford University, Stanford, California 94305, USA
[5]Department of Molecular and Cellular Physiology, Stanford University, Stanford, California 94305, USA

[‡]These authors contributed equally.

**Density and yield of SiV centers in nanopillar arrays**

We conduct a statistical study of SiV⁻ center spectra in the nanopillars to characterize the density of SiV⁻ centers and the yield of single SiV⁻ centers in nanopillar devices (Supplementary Figure 1). The number of SiV⁻ centers in each nanopillar can be identified by the number of radiative transitions in the PL spectra. The inhomogeneous distribution and the possible presence of strain-induced spectral shifting of some of the SiV⁻ centers helps identify different SiV⁻ centers in individual nanopillars. Out of the 64 nanopillars investigated within the entire array, 31.3% contain a single SiV⁻ while 6.3% contain two SiV⁻ centers. Based on the measured dimensions and the average number of SiV⁻ centers per nanopillar, we estimate the density of SiV⁻ centers in the epilayer to be $3\times10^{14}$ cm$^{-3}$.

**Second-Order Autocorrelation**

Measuring the second-order autocorrelation function g$^{(2)}$(τ) confirms the quantum nature of light emitted by the individual SiV⁻ centers in nanopillars (Supplementary Figure 2). In the representative g$^{(2)}$(τ), the SiV⁻ center is excited above saturation, and the signal is fitted with $g^{(2)}(\tau) = 1 - (1+a)e^{-|t|/\tau_1} + ae^{-|t|/\tau_2}$ convolved with the measured instrument response function (IRF), yielding $g^{(2)}(0) = 0.29$ after convolution with the IRF, and $g^{(2)}(0) = 0.04$ for ideal instrument response. The measured $g^{(2)}(0)<0.5$ supports the conclusions from the spectroscopic study that the SiV⁻ center under coherent control by the ultrafast optical pulses was indeed a single color center.

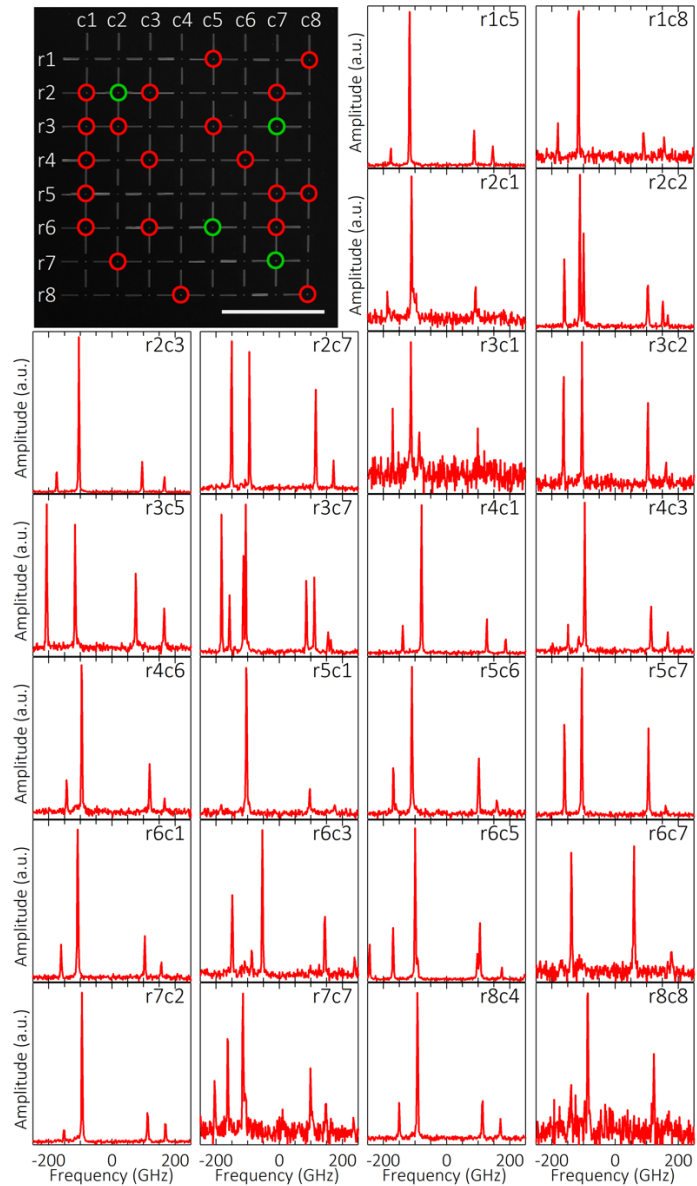

**Supplementary Figure 1.** Statistical study of the SiV⁻ center distribution in the nanopillars by PL spectroscopy. The red (green) circles indicate the nanopillars containing single (double) SiV⁻ centers, while the PL spectra of SiV⁻ centers in the nanopillars are shown in the insets.

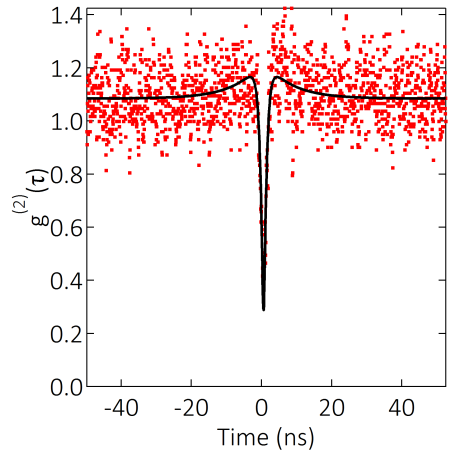

**Supplementary Figure 2.** Second-order autocorrelation function $g^{(2)}(\tau)$ of the coherently controlled SiV⁻ center, yielding $g^{(2)}(0) = 0.29$ after convolving the fitted function with the instrument response function.

18